\documentclass[twocolumn,showpacs,pra,aps]{revtex4}

\usepackage{graphicx}
\usepackage{bm}
\usepackage{psfrag}
\usepackage{amsmath}
\usepackage{amssymb}
\usepackage{latexsym}
\usepackage{exscale}
\usepackage{natbib}

\newcommand{\vect}[1]{\mathbf{#1}}
\newcommand{\sprod}{\cdot}

\newcommand{\vprod}{\times}
\newcommand{\trace}{{\rm Tr}}
\newcommand{\mi}{i}
\newcommand{\dif}{\mathrm{d}}
\newcommand{\ten}[1]{\mbox{\textbf{\textsf{#1}}}}
\renewcommand{\Im}{{\rm Im}\,}
\renewcommand{\Re}{{\rm Re}\,}
\renewcommand{\vec}[1]{\mathbf{#1}}
\newcommand{\mycomment}[1]{}
\newcommand{\Nabla}{\bm{\nabla}}

\newcommand{\D}{\mathrm{d}}

\begin{document}

\title{CASIMIR FORCE ON AMPLIFYING BODIES}

\author{Agnes Sambale, Dirk--Gunnar Welsch}

\affiliation{Theoretisch--Physikalisches Institut,
Friedrich--Schiller--Universit\"at Jena,
Max--Wien--Platz 1, D-07743 Jena, Germany}
\email{agnes.sambale@uni-jena.de}

\author{Stefan Yoshi Buhmann}
\affiliation{Quantum Optics and Laser Science, Blackett Laboratory,
Imperial College London, Prince Consort Road,
London SW7 2BW, United Kingdom}

\author{Ho Trung Dung}
\affiliation{Institute of Physics, Academy of
Sciences and Technology, 1 Mac Dinh Chi Street,
District 1, Ho Chi Minh city, Vietnam}

\date{\today}

\begin{abstract}
Based on a unified approach to macroscopic QED that allows for the
inclusion of amplification in a limited space and frequency range, we
study the Casimir force as a Lorentz force on an arbitrary partially
amplifying system of linearly locally responding (isotropic)
magnetoelectric bodies. We demonstrate that the force on a weakly
polarisable/magnetisable amplifying object in the presence of a purely
absorbing environment can be expressed as a sum over the
Casimir--Polder forces on the excited atoms inside the body. As an
example, the resonant force between a plate consisting of a dilute
gas of excited atoms and a perfect mirror is calculated.
\end{abstract}

\maketitle

\section{Introduction}
\label{sec_1}

Improved measurement methods \cite{Lamoreaux2005} and the possibility
of fabricating metamaterials \cite{Rockstuhl2007, Liu2008,Yang2008a}
and in particular, lefthanded materials (e.g.~Refs. 
\cite{Smith2000, Shelby2001, Lezec2007, Paul2008}), have motivated a
number of recent investigations into dispersion forces on micro- and
macro-objects with specially tailored magnetoelectric properties. In
this context, dispersion forces on ground-state systems such as the
Casimir--Polder (CP) interaction between a ground-state atom and a
magnetoelectric body \cite{Kampf2005} and the van der Waals (vdW)
interaction between two ground-state atoms in a magneto-electric medium
\cite{Spagnolo2007} have been studied. In both cases it was found
lefthanded medium properties, being realised in certain finite
frequency windows, are unable to noticeably affect these ground-state
forces which depend on the medium response at all frequencies in an
integral form. In the same spirit, the Casimir force between
macroscopic bodies with metamaterial properties has been studied where
possible anisotropy \cite{Rosa2008a} and lefthandedness
\cite{Yang2008} has been taken into account. It could again be shown
that the strength and the sign of the dispersion force is influenced
by the strength of the magnetic properties rather than the
lefthandedness \cite{Henkel2005, Tomas2005}.

To enhance the impact of a lefthanded magnetoelectric response, it is
natural to consider the resonant force components acting on excited
systems, which depend on single selected frequencies. The vdW
interaction has initially been studied for atoms in free space for
the cases of one ground-state and one excited atom
\cite{Power1995a,Rizzuto2004,Sherkunov2005}, two excited atoms
\cite{Power1995a} and three atoms with one of them being excited
\cite{Power1995,Passante2005}. The presence of ground-state media was
taken into account in Ref.~\cite{Sherkunov2007a}. Similarly, the 
resonant CP interaction of a macroscopic ground-state body and an
excited atom has been studied \cite{Wylie1984, Buhmann2004,
Sherkunov2007}. Applying the general results to a geometry involving
a lefthanded slab, it was noted that the inclusion of material
absorption is crucial when studying CP forces on excited atoms 
\cite{Sambale2008a, Sambale2008b}. Measurements of the CP energy of
excited atoms are typically based on spectroscopic methods 
\cite{Sandoghdar1992,Chevrollier2001,Failache2003}. 

Excitations can also be present in the electromagnetic field, with
thermal fields being an important special case. The impact of thermal
fields on the Casimir force has been subject to discussions
\cite{Ninham1998, Mostepanenko2006} and experiments \cite{Decca2005,
Klimchitskaya2005}. While the Casimir force and the CP force
\cite{McLachlan1963,Henkel2002,Gorza2006} at thermal equilibrium are 
nonresonant, integral effects just like their zero-temperature
counterparts, interesting phenomena can particularly arise for
nonequilibrium systems. For instance, the CP force on an atom near a
body held at a temperature different from that of the environment can
be attractive or repulsive depending on the temperature difference 
\cite{Antezza2005}, as has been confirmed experimentally
\cite{Harber2005, Obrecht2007}. It has further been shown that even
a ground-state atom can be subject to resonant force components when
placed in a finite-temperature environment \cite{Buhmann2008b}. The
vdW interaction between two atoms has recently even been studied in 
the presence of more general electromagnetic
fields~\cite{Sherkunov2008}. 

Although excitation has thus been included in the theories in various
forms, dispersion forces on or in the presence of amplifying media 
have not been in focus yet, although such materials are indispensable
in laser physics \cite{Scully1997} and have recently attracted
interest in the context of metamaterials
\cite{Popov2006,Stockmann2007}. In particular, they are proposed to 
lead to repulsive forces \cite{Leonhardt2007} and may therefore be
used to overcome the problem of stiction \cite{Zhao2003}. As has been
pointed out by various authors \cite{Rosa2008a, Raabe2008}, the
analysis leading to this prediction lacks a rigorous treatment of
amplification. A microscopic approach to the problem was developed in
Ref.~\cite{Sherkunov2005} where the CP potential of a ground-state
atom in front of an excited dilute gaseous medium, as well as the
Casimir interaction between two dilute samples of excited gas atoms
has been calculated. To go beyond such dilute-gas limits, an
inclusion of amplification in the quantisation scheme is necessary. To
our knowledge, the first attempt in this direction has been made in
Ref.~\cite{Jeffers1996, Matloob1997} where the light propagates
perpendicular to a dielectric amplifying slab. Within the framework of
macroscopic QED, a full picture of the quantisation of the
medium-assisted electromagnetic field in the presence of arbitrary
(absorbing or amplifying) linear, causal media has been developed very
recently~\cite{Raabe2008}. This formalism is used in the present work
to develop a consistent theory of the Casimir force on a body made of
an amplifying metamaterial which generalises microscopic results
beyond the dilute-gas limit. 

The paper is organised as follows. In Sec.~\ref{sec_2} we show how the
quantisation scheme of medium-assisted electromagnetic field should be
extended in the presence of body that is amplifying in a certain
space- and frequency regime. After calculating the Casimir force on
an arbitrary amplifying body in Sec.~\ref{volume}, we make contact to
the microscopic CP forces on the excited atoms contained inside this
body (Sec.~\ref{contact}). As an example, we consider the Casimir
force on a dilute slab of excited gas atoms. The paper ends with the
summary in Sec.~\ref{sum}. 


\section{Macroscopic quantum electrodynamics for amplifying media}
\label{sec_2}
We consider an arrangement of linear, local (isotropic)
magnetoelectric bodies some of which are (linearly) amplifying in a
limited frequency range and absorbing in the remaining frequency
range. The bodies are described by spatially varying complex electric
permittivity $\varepsilon({\bf r},\omega)$ and $\mu({\bf r},\omega)$ 
that fulfil the Kramers--Kronig relations. The electric or magnetic
response of the bodies is amplifying if $\Im\varepsilon({\bf
r},\omega) =\varepsilon_I({\bf r},\omega)<0$
or \mbox{$\Im\mu({\bf r},\omega)=\mu_I({\bf r},\omega)<0$} hold,
respectively. Note that the strength of the amplification should be
chosen such that the response to electromagnetic field is still
linear. In particular we assume that the medium-assisted field is in
an excited state where the medium is pumped in such a way that the
state of the field can be regarded as quasi-stationary (for details,
\cite{Raabe2008}). 

The quantised electric field in the presence of the partially
amplifying media can be given as the solution to the familiar
Helmholtz equation
\begin{equation}
\biggl[\bm{\nabla}\times 
\frac{1}{\mu({\bf r},\omega)}\bm{\nabla}\times
-\frac{\omega^2}{c^2}\varepsilon({\bf r},\omega)\biggr]
\hat{\underline{{\bf E}}}({\bf r},\omega) =
i\mu_0\omega\hat{\underline{{\bf j}}}_N({\bf r},\omega)
\end{equation}
according to 
\begin{equation}
\label{edef}
\hat{\underline{{\bf E}}}({\bf r},\omega)
=i\omega \mu_0\int \D ^3 r' \ten{G}({\bf r},{\bf r'},\omega)
\cdot \hat{\underline{{\bf j}}}_N({\bf r'},\omega)
\end{equation}
where the classical Green tensor obeys the differential equation
\begin{equation}
\label{e4a}
\biggl[\bm{\nabla}\times \frac{1}{\mu({\bf r},\omega)}
 \bm{\nabla}\times
 -\frac{\omega^2}{c^2}\varepsilon({\bf r},\omega)\biggr]
 \ten{G}({\bf r},{\bf r'},\omega) =
\bm{\delta}({\bf r}-{\bf r'})
\end{equation}
together with the boundary condition at infinity. 
This differential equation can be cast into its equivalent form

\begin{align}
\label{e4} 
&\biggl(\bm{\nabla}\times\bm{\nabla}\times-\frac{\omega^2}{c^2}\biggr)
\ten{G}({\bf r},
{\bf r}',
\omega)\nonumber\\
&=\bm{\delta}({\bf r}-{\bf r}')+\frac{\omega^2}{c^2}
[\varepsilon({\bf r},\omega)-1]\ten{G}({\bf r},{\bf r'},\omega)
 \nonumber\\
&\quad+\bm{\nabla}\times\biggl[1-\frac{1}{\mu({\bf r},\omega)}\biggr]
\bm{\nabla}\times\ten{G}({\bf r},{\bf r'},\omega)\nonumber\\
&=
\bm{\delta}({\bf r}-{\bf r}')
+i\mu_0\omega\int \D^3 s
\ten{Q}({\bf r}, {\bf s},
\omega)\cdot\ten{G}({\bf s}, 
{\bf r}',
\omega)
\end{align}
by introducing the conductivity tensor \cite{Kubo1998, Melrose2003}
\begin{gather}
\label{Q}
\ten{Q}({\bf r}, {\bf r'}, \omega)
=\sum_{\lambda=e,m}\ten{Q}_\lambda({\bf r}, {\bf r'}, \omega),\\
\label{Qe}
\ten{Q}_e({\bf r}, {\bf r'}, \omega)
=-i\varepsilon_0\omega[\varepsilon(
{\bf r},
\omega)-1]
\bm{\delta}({\bf r}-{\bf r'}),\\
\label{Qm}
\ten{Q}_m({\bf r}, {\bf r'}, \omega)
=-\frac{1}{i\omega\mu_0}\bm{\nabla}\!\times\!
\biggl[1\!-\!\frac{1}{\mu({\bf r},\omega)}\biggr]\!
\bm{\delta}({\bf r}\!-\!{\bf r'})\!
\times\!\overleftarrow{\bm{\nabla}}'.
\end{gather} 

It is well known that in the presence of amplification the roles of
the noise creation and noise destruction operators are to be exchanged
\cite{Scheel1998}. Hence, the noise current density reads 
\begin{multline}
\label{jn}
\hat{\underline{{\bf j}}}_N({\bf r},\omega)
=\omega\sqrt{\frac{\hbar\varepsilon_0}{\pi}|\varepsilon_I({\bf
r},\omega)|}\\
\times[\Theta[\varepsilon_I({\bf r},\omega)]\hat{{\bf f}}_e({\bf r},\omega)
+\Theta[-\varepsilon_I({\bf r},\omega)]
\hat{{\bf f}}^\dagger_e({\bf r},\omega)]\\
+\bm{\nabla}\times\sqrt{\frac{\hbar}{\pi \mu_0}
\frac{|\mu_I({\bf r},\omega)|}{|\mu({\bf r},\omega)|^2}}\\
\times[\Theta[\mu_I({\bf r},\omega)]\hat{{\bf f}}_m({\bf r},\omega)
+\Theta[-\mu_I({\bf r},\omega)]\hat{{\bf f}}^\dagger_m({\bf r},\omega)]
\end{multline}
[$\Theta$: Theta function, $\Theta(0)\equiv 1$] where the bosonic
dynamical variables $\hat{{\bf f}}_\lambda({\bf r},\omega)$ [$\lambda
= e, m$] have been introduced. They obey the commutation relations
\begin{multline}
\label{e6}
 \bigl[\hat{f}_{\lambda i}({\bf r},\omega),
 \hat{f}_{\lambda' j}({\bf r'},\omega')\bigr]=0=
 \bigl[\hat{f}^\dagger_{\lambda i}({\bf r},\omega),
 \hat{f}^\dagger_{\lambda' j}({\bf r'},\omega')\bigr]\\
 \bigl[\hat{f}_{\lambda i}({\bf r},\omega),
 \hat{f}^\dagger_{\lambda' j}({\bf r'},\omega')\bigr]=
 \delta_{\lambda\lambda'}\delta_{ij}
 \delta({\bf r}-{\bf r'})\delta(\omega-\omega'),
\end{multline}
such that the fundamental equal-time commutation relation
characteristic for the electromagnetic field holds,
\begin{equation}
 [\hat{{\bf E}}({\bf r}),\hat{{\bf B}}({\bf r'})]
 =i\hbar\varepsilon_0^{-1} \bm{\nabla}\times
 \bm{\delta}({\bf r}-{\bf r'}),
\end{equation}
where the electric and induction fields are given by Eq.~(\ref{edef})
and
\begin{equation}
\label{bdef}
\hat{\underline{{\bf B}}}({\bf r},\omega)
=\mu_0\int \D ^3 r' \bm{\nabla}\times\ten{G}({\bf r},{\bf r'},\omega)
\cdot \hat{\underline{{\bf j}}}_N({\bf r'},\omega)
\end{equation}
[$\hat{\bf O}({\bf r}) = \int _0^\infty \D \omega
\hat{\underline{\bf{O}}}({\bf r},\omega)+\rm{H.c.}$], respectively.
The vacuum state $|\bigl\{0\bigr\}\rangle$ of the body-assisted
electromagnetic field, i.e., a quasi-stationary equilibrium excited
state as introduced in the beginning of this section, is defined by
\begin{equation}
\label{e8}
\hat{{\bf f}}_\lambda ({\bf r},\omega)|\left\{0\right\}\rangle
= \bm{0} \quad \forall \lambda,{\bf r},\omega.
\end{equation}
From the quantisation procedure it follows that the Hamiltonian of the
electromagnetic field
\begin{equation}
 \hat{H}= \sum _{\lambda =e,m}\int \D ^3r \int _0^\infty\!\!\!\D
 \omega\,
 \hbar \omega \:{\rm sgn}[\kappa_\lambda
({\bf r},\omega)]
 \hat{{\bf f}}^\dagger_\lambda ({\bf r},\omega)\!\cdot\! 
 \hat{{\bf f}}_\lambda ({\bf r},\omega)
\end{equation}
[$\kappa_e=\varepsilon_I, \kappa_m=\mu_I$] generates the correct
Maxwell equations by means of the Heisenberg equations of motion.
%
%
\section{Casimir force on an amplifying body}
\label{volume}

In this section, we calculate the Casimir force on a partially
amplifying body
of volume $V$ in the presence of other bodies outside $V$. It can be
identified as the average Lorentz force \cite{Raabe2006}
\begin{multline}
\label{e20}
 {\bf F}=\int _{V
}\D ^3r \int _0^\infty
\langle \hat{\rho}({\bf r})
\hat{{\bf E}}({\bf r'})
+\hat{{\bf j}}({\bf r})\times
\hat{{\bf B}}({\bf r'})
\rangle _{{\bf r'}\to {\bf r}}
\end{multline}
where the fields are given by Eqs.~(\ref{edef}) and (\ref{bdef}) and
the internal charge and current densities read
\begin{equation}
\label{rhoint}
\underline{\hat{\rho}}({\bf r},\omega)=\frac{i\omega}{c^2}\bm{\nabla}
\cdot
\int \D ^3r' \ten{G}({\bf r}, {\bf r'}, \omega)
\cdot \underline{\hat{{\bf j}}}_N({\bf r'},\omega),
\end{equation}
\begin{equation}
\label{jint} 
\underline{\hat{{\bf j}}}({\bf r},\omega)
 =\biggl(\bm{\nabla}\times\bm{\nabla}\times-\frac{\omega^2}{c^2}
\biggr)\int \D ^3r'\ten{G}({\bf r}, {\bf r'}, \omega)\cdot
\underline{\hat{{\bf j}}}_N({\bf r'},\omega).
\end{equation}
Note that the coincidence limit ${\bf r'}\to {\bf r}$ has to be
performed in such a way that (divergent) self forces are discarded.

On using Eq.~(\ref{jn}), the commutation relations (\ref{e6}) and
Eq.~(\ref{e8}), one can easily verify  
\begin{align}
\label{jn0}
&\langle\underline{\hat{{\bf j}}}_N({\bf r},\omega)
 \underline{\hat{{\bf j}}}_N({\bf r'},\omega')\rangle
=\ten{0}
=\langle\underline{\hat{{\bf j}}}^\dagger_N({\bf r},\omega)
 \underline{\hat{{\bf j}}}^\dagger_N({\bf r'},\omega')\rangle,
\\
\label{jn1}
&\langle\underline{\hat{{\bf j}}}_N({\bf r},\omega)
 \underline{\hat{{\bf j}}}^\dagger_N({\bf r'},\omega')\rangle
 \nonumber\\
&=\frac{\hbar\omega}{\pi}\delta(\omega-\omega')
\!\!\sum_{\lambda=e,m}\!\!
\Re
\ten{Q}
_\lambda
(\vec{r},\vec{r'},\omega)\Theta[
\kappa_\lambda(\vec{r},\omega)
],
\\
\label{jn2}
&\langle\underline{\hat{{\bf j}}}_N^\dagger({\bf r},\omega)
 \underline{\hat{{\bf j}}}_N({\bf r'},\omega')\rangle
\nonumber\\
&=-\frac{\hbar\omega}{\pi}\delta(\omega-\omega')
\!\!\sum_{\lambda=e,m}\!\!
\Re
\ten{Q}
_\lambda
(\vec{r},\vec{r'},\omega)\Theta[-
\kappa_\lambda(\vec{r},\omega)]
\end{align}
for the vacuum-state expectation values. Combining these results with
the definitions~(\ref{edef}), (\ref{bdef}), (\ref{rhoint}) and
(\ref{jint}), one finds
\begin{align}
\label{rE0}
&\langle\underline{\hat{\rho}}({\bf r},\omega)
\underline{\hat{{\bf E}}}({\bf r'},\omega')\rangle
=\bm{0}=
\langle\underline{\hat{\rho}}^\dagger({\bf r},\omega)
\underline{\hat{{\bf E}}}^\dagger({\bf r'},\omega')\rangle,\\
\label{rE1}
&\langle\underline{\hat{\rho}}({\bf r},\omega)
\underline{\hat{{\bf E}}}^\dagger({\bf r'},\omega')\rangle
\nonumber\\
&=\frac{\hbar}{\pi}\frac{\omega^2}{c^2}\delta(\omega-\omega')
\mu_0\omega
\sum_{\lambda=e,m} 
\int \D^3 s   \int \D^3 s'\Theta[\kappa_\lambda(\vec{s},\omega)]
\nonumber\\
&\quad\times\bm{\nabla}\cdot\ten{G}({\bf r},{\bf s},\omega)
\cdot\Re\ten{Q}_\lambda({\bf s},{\bf s'},\omega)  \cdot
\ten{G}^\ast({\bf s'},{\bf r'},\omega),
\\
\label{rE2}
&\langle\underline{\hat{\rho}}^\dagger({\bf r},\omega)
\underline{\hat{{\bf E}}}({\bf r'},\omega')\rangle\nonumber\\
&=-\frac{\hbar}{\pi}\frac{\omega^2}{c^2}\delta(\omega-\omega')
\mu_0\omega
\!\!\sum_{\lambda=e,m} 
\int \D^3 s   \int \D^3 s'\Theta[-\kappa_\lambda(\vec{s},\omega)]
\nonumber\\
&\quad\times\bm{\nabla}\cdot\ten{G}^\ast({\bf r},{\bf
s},\omega)\cdot\Re 
\ten{Q}_\lambda({\bf s},{\bf s'},\omega)  \cdot
\ten{G}({\bf s'},{\bf r'},\omega)
\end{align}
as well as
\begin{align}
\label{jB0}
&\langle \underline{\hat{{\bf j}}}({\bf r},\omega)
 \times\underline{\hat{{\bf B}}}({\bf r'},\omega')\rangle
=\bm{0}=
\langle \underline{\hat{{\bf j}}}^\dagger({\bf r},\omega)
 \times\underline{\hat{{\bf B}}}^\dagger({\bf r'},\omega')\rangle,
 \\
\label{jB1}
&\langle \underline{\hat{{\bf j}}}({\bf r},\omega)
 \times\underline{\hat{{\bf B}}}^\dagger({\bf r'},\omega')\rangle
 \nonumber\\
&=\frac{\hbar}{\pi}\delta (\omega-\omega')\mu_0\omega
\sum_{\lambda=e,m} 
\int \D^3 s   \int \D^3 s'
\Theta[\kappa_\lambda(\vec{s},\omega)]
\nonumber\\
&\quad\times{\rm Tr}\biggl[\ten{I}\times
\biggl(\bm{\nabla}\!\times\!\bm{\nabla}\!\times\!
-\frac{ \omega^2}{c^2}\biggr)
\ten{G}({\bf r},{\bf s},\omega)\cdot
\Re \ten{Q}_\lambda({\bf s},{\bf s'},\omega)\nonumber\\  
&\quad\cdot\ten{G}^\ast({\bf s'},{\bf r'},\omega)
\times\overleftarrow{\bm{\nabla}}'\biggr],\\
\label{jB2}
&\langle \underline{\hat{{\bf j}}}^\dagger({\bf r},\omega)
 \times\underline{\hat{{\bf B}}}({\bf r'},\omega')\rangle
 \nonumber\\
&=-\frac{\hbar}{\pi}\delta (\omega-\omega')\mu_0\omega
\sum_{\lambda=e,m} 
\int \D^3 s   \int \D^3 s'
\Theta[-\kappa_\lambda(\vec{s},\omega)]
\nonumber\\
&\quad\times\!{\rm Tr}\biggl[\ten{I}\!\times\!
\biggl(\bm{\nabla}\!\times\!\bm{\nabla}\!\times
 -\frac{\omega^2}{c^2}\biggr)
\ten{G}^\ast({\bf r},{\bf s},\omega)\cdot
\Re \ten{Q}_\lambda({\bf s},{\bf s'},\omega)\nonumber\\  
&\quad\cdot\ten{G}({\bf s'},{\bf r'},\omega)
\times\overleftarrow{\bm{\nabla}}'\biggr],
\end{align}
($\ten{I}$: unit tensor, $[{\rm Tr}\ten{T}]_i=T_{kik}$), where we
have used the identity 
${\bf a}\times{\bf b}=-{\rm Tr}(\ten{I}\times{\bf a}\otimes{\bf b})$.
According to Eq.~(\ref{e20}), these results need to be added in order
to obtain the Casimir force. To that end, we make use of the identity
$\Theta[\kappa_\lambda(\vec{s},\omega)]=1
-\Theta[-\kappa_\lambda(\vec{s},\omega)]$ in Eqs.~(\ref{rE1}) and
(\ref{jB1}), combine the terms proportional to
$\Theta[-\kappa_\lambda(\vec{s},\omega)]$ with Eqs.~(\ref{rE2}) and
(\ref{jB2}) and rewrite the parts including the whole frequency
integration by means of
\begin{multline}
\mu_0\omega\int \D^3 s \int \D^3 s'\ten{G}({\bf r},{\bf s},\omega)
\cdot\Re\ten{Q}({\bf s},{\bf s'},\omega)\cdot 
\ten{G}^\ast({\bf s'},{\bf r'},\omega)\\
=\Im \ten{G}({\bf r},{\bf r'},\omega).
\end{multline}
We finally obtain the Casimir force
force
\begin{equation}
\label{F}
 {\bf F}={\bf F}^\mathrm{r}
+{\bf F}^\mathrm{nr}
\end{equation}
with
\begin{align}
\label{F1}
&{\bf F}
^\mathrm{nr}
=\frac{\hbar}{\pi}\int_{0}^\infty\dif\omega\int_{V}\dif^3r\,
 \biggl\{\frac{\omega^2}{c^2}\bm{\nabla}\sprod
 \mathrm{Im}\ten{G}(\vect{r},\vect{r'},\omega)\nonumber\\
&+\trace\biggl[\ten{I}\!\vprod\!
 \biggl(\bm{\nabla}\!\vprod\!\bm{\nabla}\!\vprod\,
 -\frac{\omega^2}{c^2}\biggr)
 \mathrm{Im}\ten{G}(\vect{r},\vect{r'},\omega)\!\vprod\!
 \overleftarrow{\bm{\nabla}}'\biggr]
 \biggr\}_{\vect{r'}\to\vect{r}}
\end{align}
and
\begin{align}
\label{F2}
&{\bf F}
^\mathrm{r}\nonumber\\
&= -\frac{2\hbar\mu_0}{\pi}\!
\int _{V}\! \D^3r\!\int_0^\infty\!\!\D \omega \omega
\!\sum_{\lambda=e,m}\!
  \int \D^3 s\! \int \D^3 s'\Theta[-\kappa_\lambda(\vec{s},\omega)]
 \nonumber\\
&\quad\times\Re\biggl\{\frac{\omega^2}{c^2}\bm{\nabla}
\cdot\ten{G}({\bf r},{\bf s},\omega)
\cdot\Re\ten{Q}_\lambda({\bf s},{\bf s'},\omega)  \cdot
\ten{G}^\ast({\bf s'},{\bf r'},\omega)\nonumber\\
&\quad+{\rm Tr}\biggl[\ten{I}\times
\biggl(\bm{\nabla}\!\times\!\bm{\nabla}\!\times\!
-\frac{ \omega^2}{c^2}\biggr)
\ten{G}({\bf r},{\bf s},\omega)\cdot
\Re \ten{Q}_\lambda({\bf s},{\bf s'},\omega)\nonumber\\  
&\quad\cdot\ten{G}^\ast({\bf s'},{\bf r'},\omega)
\times\overleftarrow{\bm{\nabla}}'\biggr]
\biggr\}_{{\bf r'}\rightarrow {\bf r}}.
\end{align}
Writing $\Im\ten{G}=(\ten{G}-\ten{G}^\ast)/(2i)$ and using the
relation $\ten{G}^\ast(\vect{r},\vect{r'},\omega)
=\ten{G}(\vect{r},\vect{r'},-\omega^\ast)$, Eq.~(\ref{F1}) can be
transformed to an integral over purely imaginary frequencies
\begin{align}
\label{F1ixi}
&{\bf F}^\mathrm{nr}
=-\frac{\hbar}{\pi}\int_{V}\dif^3r\int_{0}^\infty\dif\xi\,
 \biggl\{\frac{\xi^2}{c^2}\bm{\nabla}\sprod
 \ten{G}(\vect{r},\vect{r'},\mi\xi)
 \nonumber\\
&
-\trace\biggl[\ten{I}\vprod
\biggl(\bm{\nabla}\vprod\bm{\nabla}\vprod\,
 +\frac{\xi^2}{c^2}\biggr)
 \ten{G}(\vect{r},\vect{r'},\mi\xi)\vprod
 \overleftarrow{\bm{\nabla}}'\biggr]
 \biggr\}
 _{\vect{r'}\to\vect{r}},
\end{align}
showing that ${\bf F}^\mathrm{nr}$ is a purely nonresonant
contribution to the Casimir force. Although looking formally the same
as for purely absorbing bodies, ${\bf F}^\mathrm{nr}$
is influenced by the frequencies where the medium is
amplifying.
As evident from the factor $\Theta[-\kappa_\lambda(\vec{s},\omega)]$,
${\bf F}^\mathrm{r}$ is only nonvanishing for amplifying
bodies, it contains resonant force contributions due to emission
processes [$-\kappa_\lambda(\vec{s},\omega)$].

%
\section{Contact to Casimir--Polder forces}
\label{contact}
In order to interpret the two contributions ${\bf F}^\mathrm{nr}$ and
${\bf F}^\mathrm{r}$ as the Casimir force on an amplifying body, it
is instructive to establish the relation of the Casimir
force~(\ref{F})-(\ref{F2}) to the more well-known CP forces on
excited atoms. To that end, we consider the Casimir force on an
optically dilute amplifying body of volume $V$ placed in a free-space
region in an environment of purely absorbing bodies
(cf.~Fig.~\ref{Fig2} for an example). We follow the procedure
outlined in Ref.~\cite{Raabe2006} for an absorbing dielectric body,
starting with~${\bf F}^{\rm nr}$. First, we express the nonresonant
force~(\ref{F1}) in terms of the electric and magnetic susceptibilities
$\varepsilon({\bf r},\omega)-1$ and $1-1/\mu({\bf r},\omega)$ (${\bf
r}\in V$) of the body by invoking the relations
\begin{align}
\label{eq31} 
&\biggl(\bm{\nabla}\times\bm{\nabla}\times-\frac{\omega^2}{c^2}\biggr)
\Im\ten{G}({\bf r},{\bf r}',\omega)\nonumber\\
&=\frac{\omega^2}{c^2}
\Im\{[\varepsilon({\bf r},\omega)-1]\ten{G}({\bf r},{\bf r'},\omega)\}
 \nonumber\\
&\quad+\bm{\nabla}\times\Im\biggl\{\biggl[1-\frac{1}{\mu({\bf
r},\omega)}\biggr]
\bm{\nabla}\times\ten{G}({\bf r},{\bf r'},\omega)\biggr\},\\
&\bm{\nabla}\cdot\Im\ten{G}({\bf r},{\bf r}',\omega)\nonumber\\
&=-\bm{\nabla}\cdot\Im
\{[\varepsilon({\bf r},\omega)-1]\ten{G}({\bf r},{\bf r'},\omega)\},
\end{align}
which follow from the differential equation~(\ref{e4}). Expanding the
results with the aid of the identities
\begin{align}
\label{cross1}
&{\rm Tr}\bigl[\ten{I}\times\ten{G}({\bf r},{\bf r'},\omega)
\times\overleftarrow{\bm{\nabla}}'\bigr]\nonumber\\
&\qquad=\bm{\nabla}'{\rm Tr}\ten{G}({\bf r},{\bf r'},\omega)
 -\bm{\nabla}'\cdot\ten{G}({\bf r},{\bf r'},\omega),\\
\label{cross2}
&{\rm Tr}\bigl\{\ten{I}\times[\bm{\nabla}\times
\ten{H}({\bf r},{\bf r'},\omega)]\bigr\}\nonumber\\
&\qquad=\ten{H}({\bf r},{\bf r'},\omega)
 \cdot\overleftarrow{\bm{\nabla}}'
 -\bm{\nabla}{\rm Tr}\ten{H}({\bf r},{\bf r'},\omega)
\end{align} 
[$\ten{H}({\bf r},{\bf r'},\omega)=\bm{\nabla}\times
\ten{G}({\bf r},{\bf r'},\omega)\times\overleftarrow{\bm{\nabla}}'$]
and exploiting the fact that terms involving a total divergence can
be converted to vanishing surface integrals for a body in free space,
one obtains
\begin{align}
\label{F1dilute}
&{\bf F}^\mathrm{nr}
=\frac{\hbar}{2\pi}\int_{0}^\infty\dif\omega\int_{V}\dif^3r\,
\nonumber\\
&\times\biggl(\frac{\omega^2}{c^2}
 \Im\{[\varepsilon({\bf r},\omega)-1]\bm{\nabla}
 {\rm Tr}
 [\ten{G}(\vect{r},\vect{r'},\omega)]_{\vect{r'}\to\vect{r}}
 \}\nonumber\\
&-\bm{\nabla}\Im\biggl\{
 \biggl[1-\frac{1}{\mu({\bf r},\omega)}\biggr]
 \trace\bigl[\bm{\nabla}\vprod
 \ten{G}(\vect{r},\vect{r'},\omega)\vprod
 \overleftarrow{\bm{\nabla}}'\bigr]_{\vect{r'}\to\vect{r}}
 \biggr).
\end{align}
Note that $[\bm{\nabla}'{\rm Tr}
 \ten{G}(\vect{r},\vect{r'},\omega)]_{\vect{r'}\to\vect{r}}
=\frac{1}{2}\bm{\nabla}{\rm Tr}
 [\ten{G}(\vect{r},\vect{r'},\omega)]_{\vect{r'}\to\vect{r}}$
due to the symmetry $\ten{G}(\vect{r}',\vect{r},\omega)
=\ten{G}^\mathrm{T}(\vect{r},\vect{r}',\omega)$ of the Green tensor
\cite{Raabe2006}. For a homogeneous body, the coincidence limit can be
performed by simply replacing the Green tensor with its scattering
part $\ten{G}^{(1)}$ (see the discussion in Ref.~\cite{Raabe2006}).
Rewriting the result as an integral over purely imaginary frequencies
[cf.\ the remarks above Eq.~(\ref{F1ixi})], one obtains
\begin{align}
\label{F1diluteixi}
&{\bf F}^\mathrm{nr}\nonumber\\
&=-\frac{\hbar}{2\pi}\int_{0}^\infty\!\!\!\dif\xi\int_{V}\dif^3r\,
 \biggl(\frac{\xi^2}{c^2}
 [\varepsilon({\bf r},i\xi)-1]\bm{\nabla}
 {\rm Tr}
 \ten{G}^{(1)}(\vect{r},\vect{r},i\xi)
 \nonumber\\
&\quad+\biggl[1-\frac{1}{\mu({\bf r},i\xi)}\biggr]
 \bm{\nabla}\trace\bigl[\bm{\nabla}\!\vprod\!
 \ten{G}^{(1)}(\vect{r},\vect{r},i\xi)\!\vprod\!
 \overleftarrow{\bm{\nabla}}'\bigr]\biggr).
\end{align}

Next, we make use of the fact that the amplifying body is assumed to
be optically dilute by expanding the result~(\ref{F1diluteixi}) to
leading, linear order in the susceptibilities 
$\varepsilon({\bf r},\omega)-1$ and $1-1/\mu({\bf r},\omega)$ 
(${\bf r}\in V$) of the amplifying body. Since these susceptibilities
already explicitly appear as factors in the above expression, the
Green tensors have to be expanded to zeroth order in these functions.
In other words, we have to replace $\ten{G}$ with the Green tensor
$\overline{\ten{G}}$ of the system in the absence of the amplifying
body, which is the solution to the Helmholtz equation~(\ref{e4a}) with
\begin{equation}
\label{epsilonmubar}
\overline{\varepsilon}({\bf r},\omega),\overline{\mu}({\bf r},\omega)
=\begin{cases}
\varepsilon({\bf r},\omega),\mu({\bf r},\omega)\quad\mbox{for }
 {\bf r}\notin V,\\
1 \quad\mbox{for }{\bf r}\in V.
\end{cases}
\end{equation}
in place of $\varepsilon({\bf r},\omega)$ and $\mu({\bf r},\omega)$.

And finally, we assume that the amplifying body consists of a gas of
isotropic atoms in an excited state $|n\rangle$ with polarisability
\begin{equation}
\label{alpha}
\alpha_n(\omega)=\lim _{\epsilon\rightarrow 0} \frac{1}{3\hbar}
\sum_k\biggl[
\frac{|\mathbf{d}_{nk}|^2}{\omega+\omega_{kn}+i\epsilon}
-\frac{|\mathbf{d}_{nk}|^2}{\omega-\omega_{kn}+i\epsilon}\biggr]
\end{equation}
($\omega_{kn}$: transition frequencies, $\mathbf{d}_{nk}$: electric
dipole matrix elements) and magnetisability
\begin{equation}
\label{beta}
\beta_n(\omega)=\lim _{\epsilon\rightarrow 0} \frac{1}{3\hbar}
\sum_k\biggl[
\frac{|\mathbf{m}_{nk}|^2}{\omega+\omega_{kn}+i\epsilon}
-\frac{|\mathbf{m}_{nk}|^2}{\omega-\omega_{kn}+i\epsilon}\biggr]
\end{equation}
($\mathbf{m}_{nk}$: magnetic dipole matrix elements). Relating
electric and magnetic susceptibilities of the body with the atomic
polarisability and magnetisability via the linearised
Clausius--Mossotti laws
\begin{equation}
\label{CM}
 \varepsilon(\omega)-1=
 \frac{\eta\alpha_n(\omega)}{\varepsilon_0},
 \qquad
 1-\frac{1}{\mu(\omega)}=\mu_0\eta\beta_n(\omega),
\end{equation}
($\eta$: atomic number density), one obtains
\begin{equation}
\label{relationr}
{\bf F}^{\rm nr}=\int\D^3r\eta\bm{\nabla}U_n^{\rm nr}({\bf r})
\end{equation}
where
\begin{align}
\label{F1od}
U_n^{\rm nr}({\bf r})
=&
\frac{\hbar\mu_0}{2\pi}
\int _0^\infty \D\xi\Bigl\{
\xi^2\alpha_n(i\xi)
{\rm Tr}\overline{\ten{G}}^{(1)}(\vec{r},\vec{r},i\xi)\nonumber\\
&+\beta_n(i\xi)
{\rm Tr}\bigl[\Nabla\times
\overline{\ten{G}}^{(1)}(\vec{r},\vec{r},i\xi)\times
\overleftarrow{{\Nabla}}'\bigr]\Bigr\}.
\end{align}
The nonresonant Casimir force on a optically dilute amplifying body is
hence a summation over the respective nonresonant CP forces on the
excited atoms the body consists of, where the nonresonant CP
potential~(\ref{F1od}) generalises well-known expressions for
polarisable atoms \cite{Buhmann2004pertur, Buhmann2004}
to the case of magnetoelectric atoms. The nonresonant potential of
excited atoms obtained here is invariant with respect to a duality
transformation, i.e., a simultaneous global exchange
$\alpha_n\leftrightarrow\beta_n/c^2$ and
$\varepsilon\leftrightarrow\mu$, as can generally be expected for
atoms in free space \cite{Buhmann2008a}. Note that there is one
important difference to the case of the force on an absorbing object
which consists of ground-state atoms: While for ground-state atoms all
of the frequencies $\omega_{kn}$ in Eqs.~(\ref{alpha}) and
(\ref{beta}) are positive so that all (virtual) transitions contribute
to the nonresonant CP potential with the same sign, upward as well as
downward transitions are possible for excited atoms, so that positive
and negative $\omega_{kn}$ occur. In particular for a two-level atom,
the nonresonant CP force for the atom in its excited state is exactly
opposite to the respective ground-state force.

Let us next consider the resonant Casimir force ${\bf F}^\mathrm{r}$,
which is only present for an amplifying body, by following
essentially the same steps as for the nonresonant force. Substituting
Eqs.~(\ref{Qe}) and (\ref{Qm}) in to Eq.~(\ref{F2}), we obtain
\begin{widetext}
\begin{eqnarray}
\label{F2dilute}
{\bf F}^\mathrm{r}
&=& -\frac{2\hbar}{\pi}\!
\int _{V}\! \D^3r\!\int_0^\infty\!\!\D \omega 
 \int \D^3 s\, 
 \biggl(\Theta[-\varepsilon_I(\vec{s},\omega)]
 \frac{\omega^2}{c^2}
 \biggl\{\frac{\omega^2}{c^2}\bm{\nabla}
\cdot\Re\bigl[\ten{G}({\bf r},{\bf s},\omega)
\cdot\varepsilon_I({\bf s},\omega)  \cdot
\ten{G}^\ast({\bf s},{\bf r'},\omega)\bigr]\nonumber\\
&&+{\rm Tr}\,\Re\biggl[\ten{I}\times
\biggl(\bm{\nabla}\!\times\!\bm{\nabla}\!\times\!
-\frac{ \omega^2}{c^2}\biggr)
\ten{G}({\bf r},{\bf s},\omega)
\varepsilon_I({\bf s},\omega) 
\cdot\ten{G}^\ast({\bf s},{\bf r'},\omega)
\times\overleftarrow{\bm{\nabla}}'\biggr]
\biggr\}\nonumber\\
&&+\Theta[-\mu_I(\vec{s},\omega)]\biggl\{
\frac{\omega^2}{c^2}\bm{\nabla}
\cdot\Re\bigl[\ten{G}({\bf r},{\bf s},\omega)
\cdot\Nabla_s\times
\frac{\mu_I({\bf s},\omega)}{|\mu({\bf s},\omega)|^2}
\Nabla_s\times
\ten{G}^\ast({\bf s},{\bf r'},\omega)\bigr]
\nonumber\\
&&\times{\rm Tr}\,\Re\biggl[\ten{I}\times
\biggl(\bm{\nabla}\!\times\!\bm{\nabla}\!\times\!
-\frac{ \omega^2}{c^2}\biggr)
\ten{G}({\bf r},{\bf s},\omega)
\cdot\Nabla_s\times
\frac{\mu_I({\bf s},\omega)}{|\mu({\bf s},\omega)|^2}
\Nabla_s\times
\ten{G}^\ast({\bf s},{\bf r'},\omega)
\times\overleftarrow{\bm{\nabla}}'\biggr]
\biggr\}
 \biggr)_{{\bf r'}\rightarrow {\bf r}},
\end{eqnarray}
\end{widetext}
where in contrast to the respective nonresonant result~(\ref{F1}), the
susceptibilities of the amplifying body are already explicitly present
at this stage. 

A linear approximation in these susceptilities can hence be obtained
by using the zero-order identities
\begin{gather}
\label{eq44} 
\biggl(\bm{\nabla}\times\bm{\nabla}\times-\frac{\omega^2}{c^2}\biggr)
\ten{G}({\bf r},{\bf r}',\omega)
=\bm{\delta}({\bf r}-{\bf r}'),\\
\label{eq45}
\frac{\omega^2}{c^2}\Nabla\cdot\ten{G}({\bf r},{\bf r}',\omega)
=-\Nabla\delta({\bf r}-{\bf r}')
\end{gather}
[cf.\ Eq.~(\ref{e4})] and replacing $\ten{G}^\ast$ with
$\overline{\ten{G}}^\ast$. Expanding the result with the aid of
Eqs.~(\ref{cross1}) and (\ref{cross2}) and discarding terms involving
total divergences for a body in free space, one finds
\begin{align}
\label{Fod4}
{\bf F}^\mathrm{r}
=&-\frac{\hbar}{\pi}\int_{0}^\infty\dif\omega\int_{V}\dif^3r\,
 \biggl\{\Theta[-\varepsilon_I({\bf r},\omega)]
 \frac{\omega^2}{c^2}
 \varepsilon_I({\bf r},\omega)\nonumber\\
&\times\bm{\nabla}
 {\rm Tr}\,\Re
 \ten{G}^{(1)}(\vect{r},\vect{r},\omega)
 -\Theta[-\mu_I({\bf r},\omega)]\nonumber\\
&\times
\frac{\mu_I({\bf r},\omega)}{|\mu({\bf r},\omega)|^2}
 \bm{\nabla}\trace\bigl[\bm{\nabla}\!\vprod\!
 \Re\ten{G}^{(1)}(\vect{r},\vect{r},\omega)\!\vprod\!
 \overleftarrow{\bm{\nabla}}'\bigr]\biggr\},
\end{align}
where we have again assumed the body to be homogeneous and performed
the coincidence limit by replacing the Green tensor with its
scattering part.

Relating $\varepsilon_I$ and $\mu_I$ to the polarisability and
magnetisability of the atoms by means of the Clausius--Mossotti
relation~(\ref{CM}) we finally obtain
\begin{equation}
\label{Fod5}
 {\bf F}
^\mathrm{r} 
=\int \D ^3 r
\eta\bm{\nabla}U_n^{\rm r}({\bf r})
\end{equation}
where
\begin{align}
\label{eq48}
U_n^\mathrm{r}({\bf r})=&
\frac{\hbar\mu_0}{\pi}\int_{0}^\infty\dif\omega
 \biggl\{\Theta[-\Im\alpha_n(\omega)]
 \Im\alpha_n(\omega)\nonumber\\
&\times\omega^2
 {\rm Tr}\,\Re
 \ten{G}^{(1)}(\vect{r},\vect{r},\omega)
 -\Theta[-\Im\beta_n(\omega)]\nonumber\\
&\times
 \Im\beta_n(\omega)
 \trace\bigl[\bm{\nabla}\!\vprod\!
 \Re\ten{G}^{(1)}(\vect{r},\vect{r},\omega)\!\vprod\!
 \overleftarrow{\bm{\nabla}}'\bigr]\biggr\}
\end{align}
is the resonant part of the CP potential of the excited atoms
contained in the body. By using the relations
\begin{eqnarray}
\label{Imalpha}
 \alpha_I(\omega)=\frac{\pi}{3\hbar}\sum_k |{\bf d}_{nk}|^2 
       [\delta(\omega+\omega_{nk})-\delta(\omega-\omega_{nk})],\\
\label{Imbeta}
 \beta_I(\omega)=\frac{\pi}{3\hbar}\sum_k |{\bf m}_{nk}|^2 
       [\delta(\omega+\omega_{nk})-\delta(\omega-\omega_{nk})],
\end{eqnarray}
which follow from the definitions~(\ref{alpha}) and (\ref{beta}) by
means of the identity $\lim_{\epsilon\to
0}1/(x+i\epsilon)=\mathcal{P}/x-i\pi\delta(x)$ ($\mathcal{P}$:
principal value), the resonant CP potential can be written in the
more familiar form
\begin{align}
\label{Fod6}
&U_n^\mathrm{r}({\bf r})
=-\frac{\mu_0}{3}\sum _k
\Theta(\omega_{nk})\Bigl\{\omega_{nk}^2|{\bf d}_{nk}|^2
{\rm Tr Re} \mathbf{G}^{(1)}({\bf r},{\bf r}, \omega_{nk})
 \nonumber\\
&\quad-|{\bf m}_{nk}|^2\trace\bigl[\bm{\nabla}\!\vprod\!
 \Re\ten{G}^{(1)}(\vect{r},\vect{r},\omega)\!\vprod\!
 \overleftarrow{\bm{\nabla}}'\bigr]\Bigr\},
\end{align}
which generalises previous results for purely electric atoms
\cite{Buhmann2004} to the magnetoelectric case. The resonant part of
the CP potential is obviously associated with real, energy-conserving
transitions of the excited atom to lower states, it dominates over the
nonresonant part of the potential, in general. As expected, the
resonant part of the CP potential of an excited atom in free space is
duality-invariant, just like the nonresonant part \cite{Buhmann2008a}.

Combining our results~(\ref{relationr}) and (\ref{Fod5}) in
accordance with Eq.~(\ref{F}), we have shown that the Casimir force on
a optically dilute, homogeneous, amplifying magnetoelectric body
is the sum of the CP forces on the excited atoms contained in it,
\begin{equation}
\label{eq52}
 {\bf F}=-\int \D ^3 r
\eta
\Nabla U_n({\bf r}).
\end{equation}
This main result of this section generalises similar findings for
purely absorbing bodies (consisting of ground-state atoms)
\cite{Tomas2005b, Raabe2006, Buhmann2006a, Buhmann2006b} to the
amplifying case. In addition, our calculation has rendered explicit
expressions for the free-space CP potential of excited magnetoelectric
atoms in the presence of an arbitrary arrangement of absorbing bodies,
\begin{equation}
\label{eq53}
U_n({\bf r})=U_n^\mathrm{nr}({\bf r})+U_n^\mathrm{r}({\bf r}),
\end{equation}
with $U_n^\mathrm{nr}$ and $U_n^\mathrm{r}$ being given by
Eqs.~(\ref{F1od}) and (\ref{Fod6}), respectively. In this dilute gas
limit, the nonresonant and resonant components of the Casimir
force~(\ref{F1}) and (\ref{F2}) are directly related to the
respective CP-potential components which in turn are associated with
virtual and real transitions of the atoms. Our considerations contain
the case of a purely absorbing body consisting of ground-state
atoms~\cite{Tomas1995} as a special case. The most important
difference between forces on ground-state versus excited atoms is the
contribution from possible real transitions only present for excited
atoms which manifests itself as the resonant contribution~(\ref{F2})
of the Casimir force. Note that the established direct relation
between Casimir forces and single-atom CP forces is only valid for
dilute bodies, while for bodies with stronger magnetoelectric
properties, many-atom interaction begin to play a role and lead to a
breakdown of additivity (cf., e.g.,Refs.~\cite{Buhmann2006a,
Buhmann2006b}). 
%
%
\section{Example: Force on a plate of excited atoms near a perfect
mirror}
\label{gas}
\begin{figure}[t]
\begin{center}
\includegraphics[width=\linewidth]{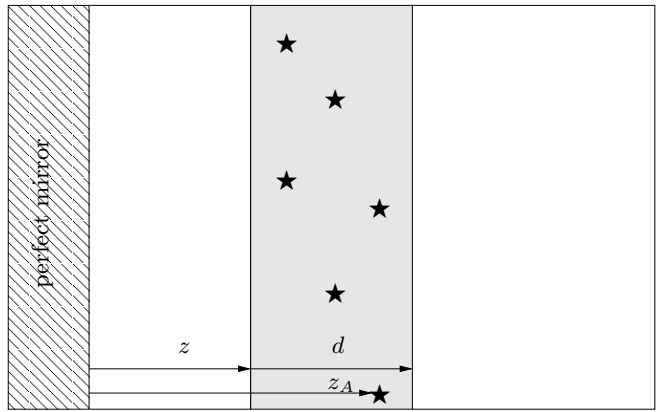}
\end{center}
\caption{
Amplifying slab consisting of a gas of excited atoms in front of
perfect
mirror.}
\label{Fig2}
\end{figure}

\begin{figure}[t]
\begin{center}
\includegraphics[width=\linewidth]{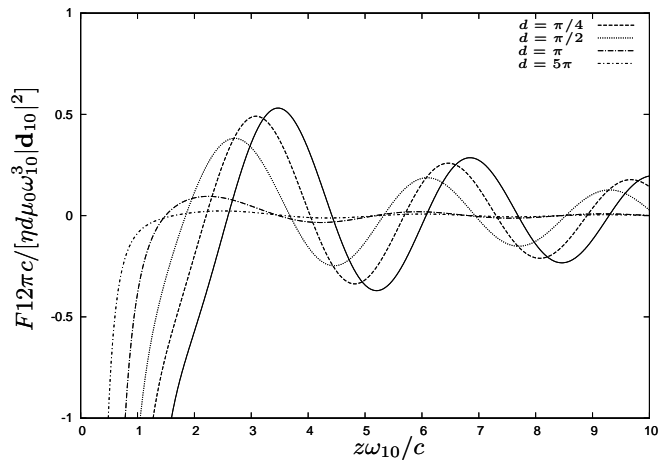}
\end{center}
\caption{Resonant Casimir force (per thickness $d$) between a planar
sample of excited gas atoms and a perfect mirror plotted vs.~the
plate--mirror distance. The atomic dipole moments are oriented
parallel to the surface. The solid line shows the resonant CP force on
each excited atom.}
\label{Fig3}
\end{figure}

Let us apply our theory to a plate (thickness $d$) consisting of a
gas of excited, purely electric (two-level) gas atoms which is
situated at a distance $z$ from a perfectly reflecting
mirror (see Fig.~\ref{Fig2}).
%
The associated Green tensor reads \cite{Sambale2008a}
\begin{align}
\label{gr1}
G_{xx}^{(1)}(z_A,z_A,\omega_{10})&=G_{yy}^{(1)}(z_A,z_A,\omega_{10})
\nonumber\\
&=\frac{\omega_{10} e^{i\tilde{z}}}{4\pi c\tilde{z}^3}
\left(1-i\tilde{z}-\tilde{z}^2\right),
\end{align}
\begin{equation}
\label{gr2}
G_{zz}^{(1)}(z_A,z_A,\omega_{10})
=\frac{\omega_{10}e^{i\tilde{z}}}{2\pi c \tilde{z}^3}
\left(1-i\tilde{z}\right)
\end{equation}
[$\tilde{z}=2\omega_{10}z_A/c$] were $z_A$ denotes the atom--mirror
distance. We
will only consider the dominant
resonant component of the Casimir force,
$\vec{F}\approx \vec{F}_r$. According to Eq.~(\ref{Fod5}) together
with (\ref{Fod6}) the Casimir force per unit area on the weakly
polarisable plate is given by
\begin{align}
 \vec{F}(z)=&\frac{\mu_0}{3}\eta\omega_{10}^2 |\vec{d}_{10}|^2
\int_{z}^{z+d}\D
z_A 
\frac{\partial}{\partial{z_A}}\Re G_{ii}^{(1)}
(\vec{r}_A,\vec{r}_A,\omega)\vec{e}_z\\
\equiv &\frac{\mu_0}{3}\eta\omega_{10}^2 |\vec{d}_{10}|^2
H(\vec{z})\vec{e}_z
\end{align}
[$z$: plate--mirror distance, $\eta$: density of atoms in the plate]
where
\begin{multline}
\label{equ21}
H(z)=
\frac{\omega_{10}}{4\pi c \tilde{z}^3}
 \left[(2-\tilde{z}^2)\cos(\tilde{z})+2\tilde{z}\sin(\tilde{z})
\right]_{z_A=z}^{z_A=z+d}.
\end{multline}
Fig.~(\ref{Fig3}) shows the (dimensionless) Casimir force 
per thickness $d$ of the plate as a function of the atom--plate
separation. For reference, we have also displayed the resonant part of
the CP force on the individual atoms contained in the plate. It is
seen that the Casimir force on the amplifying plate shows an
attractive behaviour in the short-distance regime while for big
plate--mirror separations an oscillating behavior can be found. This
is a direct consequence of the respective behaviour of the CP forces
on the atoms contained in the plate. The amplitude of the oscillations
decreases with increasing thickness of the plate, since the integrated
Casimir force per plate thickness is a spatial average of the
oscillating CP forces over the plate thickness. The occurrence of
oscillations can be regarded as the typical impact of amplification on
the Casimir force.
%
%
%
\section{Summary}
\label{sum}

Starting from the quantisation of the electromagnetic field in the
presence of arbitrary linearly and locally responding media we have
derived formulas for Casimir forces in a partially amplifying
arrangement of (isotropic) magnetoelectric bodies of arbitrary shapes.
Treating the Casimir force in the Lorentz force approach we have 
found that it can be decomposed into two parts; one that looks
formally the same as in the case of purely absorbing body and can be 
regarded as an off-resonant force component due to the integration 
over the full frequency range and, a second, resonant part of the
Casimir force that is a direct consequence of the amplification in the
system and contains an integration over the frequencies where the 
imaginary part of the electric permittivity and/or (para)magnetic
permeability is negative. These off-resonant and resonant parts of the
Casimir force have a simple physical interpretation in the case of a
partially amplifying object consisting of weak
polarisable/magnetisable material in a purely absorbing environment.
We have demonstrated that in this approximation the Casimir force can
be written as a summation of pairwise CP forces on the excited atoms
the body consists of; where the off-resonant/resonant contribution
to the Casimir force could be related to the off-resonant/resonant
CP force. These in turn are associated with virtual and real
transitions of the excited atoms. As an example, we have studied the
Casimir force between a weakly polarisable plate of excited gas atoms
and a perfectly reflecting mirror. We have found attraction to the
surface for small mirror--plate separation while the retarded regime
is dominated by an oscillating behaviour as a consequence of the
spatial average of the respective CP forces of the excited gas atoms. 

The theory can be expanded to allow for linearly but non-locally
responding anisotropic media. It can be used to evaluate the Casimir
force beyond linear order in the susceptibilities.


\acknowledgments

The work was supported by Deutsche Forschungsgemeinschaft. We
acknowledge funding from the Alexander von Humboldt Foundation
(S.~Y.~B. and H.~T.~D.) and the Vietnamese Basic Research Program
(H.~T.~D.). A.S. acknowledges fruitful discussions with C. Raabe.

\bibliographystyle{apsrev}

\begin{thebibliography}{58}
\expandafter\ifx\csname natexlab\endcsname\relax\def\natexlab#1{#1}\fi
\expandafter\ifx\csname bibnamefont\endcsname\relax
  \def\bibnamefont#1{#1}\fi
\expandafter\ifx\csname bibfnamefont\endcsname\relax
  \def\bibfnamefont#1{#1}\fi
\expandafter\ifx\csname citenamefont\endcsname\relax
  \def\citenamefont#1{#1}\fi
\expandafter\ifx\csname url\endcsname\relax
  \def\url#1{\texttt{#1}}\fi
\expandafter\ifx\csname urlprefix\endcsname\relax\def\urlprefix{URL }\fi
\providecommand{\bibinfo}[2]{#2}
\providecommand{\eprint}[2][]{\url{#2}}

\bibitem[{\citenamefont{Lamoreaux}(2005)}]{Lamoreaux2005}
\bibinfo{author}{\bibfnamefont{S.~K.} \bibnamefont{Lamoreaux}},
  \bibinfo{journal}{{R}ep. {P}rog. {P}hys.} \textbf{\bibinfo{volume}{68}},
  \bibinfo{pages}{201} (\bibinfo{year}{2005}).

\bibitem[{\citenamefont{Rockstuhl et~al.}(2007)\citenamefont{Rockstuhl,
  Lederer, Etrich, Pertsch, and Scharf}}]{Rockstuhl2007}
\bibinfo{author}{\bibfnamefont{C.}~\bibnamefont{Rockstuhl}},
  \bibinfo{author}{\bibfnamefont{F.}~\bibnamefont{Lederer}},
  \bibinfo{author}{\bibfnamefont{C.}~\bibnamefont{Etrich}},
  \bibinfo{author}{\bibfnamefont{T.}~\bibnamefont{Pertsch}}, \bibnamefont{and}
  \bibinfo{author}{\bibfnamefont{T.}~\bibnamefont{Scharf}},
  \bibinfo{journal}{{P}hys. {R}ev. {L}ett.} \textbf{\bibinfo{volume}{99}},
  \bibinfo{pages}{017401} (\bibinfo{year}{2007}).

\bibitem[{\citenamefont{Liu et~al.}(2008)\citenamefont{Liu, Guo, Fu, Kaiser,
  Schweizer, and Giessen}}]{Liu2008}
\bibinfo{author}{\bibfnamefont{N.}~\bibnamefont{Liu}},
  \bibinfo{author}{\bibfnamefont{H.}~\bibnamefont{Guo}},
  \bibinfo{author}{\bibfnamefont{L.}~\bibnamefont{Fu}},
  \bibinfo{author}{\bibfnamefont{S.}~\bibnamefont{Kaiser}},
  \bibinfo{author}{\bibfnamefont{H.}~\bibnamefont{Schweizer}},
  \bibnamefont{and} \bibinfo{author}{\bibfnamefont{H.}~\bibnamefont{Giessen}},
  \bibinfo{journal}{{N}ature {M}at.} \textbf{\bibinfo{volume}{7}},
  \bibinfo{pages}{31} (\bibinfo{year}{2008}).

\bibitem[{\citenamefont{Yang and Li}(2008)}]{Yang2008a}
\bibinfo{author}{\bibfnamefont{Y.}~\bibnamefont{Yang}} \bibnamefont{and}
  \bibinfo{author}{\bibfnamefont{G.}~\bibnamefont{Li}, \bibfnamefont{Q.~Wang}}
  (\bibinfo{year}{2008}).

\bibitem[{\citenamefont{Smith et~al.}(2000)\citenamefont{Smith, Padilla, Vier,
  Nemat-Nasser, and Schultz}}]{Smith2000}
\bibinfo{author}{\bibfnamefont{D.~R.} \bibnamefont{Smith}},
  \bibinfo{author}{\bibfnamefont{W.~J.} \bibnamefont{Padilla}},
  \bibinfo{author}{\bibfnamefont{D.~C.} \bibnamefont{Vier}},
  \bibinfo{author}{\bibfnamefont{S.~C.} \bibnamefont{Nemat-Nasser}},
  \bibnamefont{and} \bibinfo{author}{\bibfnamefont{S.}~\bibnamefont{Schultz}},
  \bibinfo{journal}{Phys. {R}ev. {L}ett.} \textbf{\bibinfo{volume}{84}},
  \bibinfo{pages}{4184} (\bibinfo{year}{2000}).

\bibitem[{\citenamefont{Shelby et~al.}(2001)\citenamefont{Shelby, Smith,
  Nemat-Nasser, and Schultz}}]{Shelby2001}
\bibinfo{author}{\bibfnamefont{R.~A.} \bibnamefont{Shelby}},
  \bibinfo{author}{\bibfnamefont{D.~R.} \bibnamefont{Smith}},
  \bibinfo{author}{\bibfnamefont{S.~C.} \bibnamefont{Nemat-Nasser}},
  \bibnamefont{and} \bibinfo{author}{\bibfnamefont{S.}~\bibnamefont{Schultz}},
  \bibinfo{journal}{App. {P}hys. {L}ett.} \textbf{\bibinfo{volume}{78}},
  \bibinfo{pages}{489} (\bibinfo{year}{2001}).

\bibitem[{\citenamefont{Lezec et~al.}(2007)\citenamefont{Lezec, Dionne, and
  Atwater}}]{Lezec2007}
\bibinfo{author}{\bibfnamefont{H.~J.} \bibnamefont{Lezec}},
  \bibinfo{author}{\bibfnamefont{J.~A.} \bibnamefont{Dionne}},
  \bibnamefont{and} \bibinfo{author}{\bibfnamefont{H.~A.}
  \bibnamefont{Atwater}}, \bibinfo{journal}{Sci.}
  \textbf{\bibinfo{volume}{316}}, \bibinfo{pages}{430} (\bibinfo{year}{2007}).

\bibitem[{\citenamefont{Paul et~al.}(2008)\citenamefont{Paul, Imhof, Reinhard,
  Zengerle, and Beigang}}]{Paul2008}
\bibinfo{author}{\bibfnamefont{O.}~\bibnamefont{Paul}},
  \bibinfo{author}{\bibfnamefont{C.}~\bibnamefont{Imhof}},
  \bibinfo{author}{\bibfnamefont{B.}~\bibnamefont{Reinhard}},
  \bibinfo{author}{\bibfnamefont{R.}~\bibnamefont{Zengerle}}, \bibnamefont{and}
  \bibinfo{author}{\bibfnamefont{R.}~\bibnamefont{Beigang}},
  \bibinfo{journal}{{O}pt. {E}xpr.} \textbf{\bibinfo{volume}{16}},
  \bibinfo{pages}{6736} (\bibinfo{year}{2008}).

\bibitem[{\citenamefont{Buhmann et~al.}(2005)\citenamefont{Buhmann, Welsch, and
  Kampf}}]{Kampf2005}
\bibinfo{author}{\bibfnamefont{S.~Y.} \bibnamefont{Buhmann}},
  \bibinfo{author}{\bibfnamefont{D.-G.} \bibnamefont{Welsch}},
  \bibnamefont{and} \bibinfo{author}{\bibfnamefont{T.}~\bibnamefont{Kampf}},
  \bibinfo{journal}{Phys. {R}ev. {A}} \textbf{\bibinfo{volume}{72}},
  \bibinfo{pages}{032112} (\bibinfo{year}{2005}).

\bibitem[{\citenamefont{Spagnolo et~al.}(2007)\citenamefont{Spagnolo, Dalvit,
  and Milonni}}]{Spagnolo2007}
\bibinfo{author}{\bibfnamefont{S.}~\bibnamefont{Spagnolo}},
  \bibinfo{author}{\bibfnamefont{D.~A.~R.} \bibnamefont{Dalvit}},
  \bibnamefont{and} \bibinfo{author}{\bibfnamefont{P.~W.}
  \bibnamefont{Milonni}}, \bibinfo{journal}{Phys. {R}ev. A}
  \textbf{\bibinfo{volume}{75}}, \bibinfo{pages}{052117}
  (\bibinfo{year}{2007}).

\bibitem[{\citenamefont{Rosa et~al.}(2008)\citenamefont{Rosa, Dalvit, and
  Milonni}}]{Rosa2008a}
\bibinfo{author}{\bibfnamefont{F.~S.~S.} \bibnamefont{Rosa}},
  \bibinfo{author}{\bibfnamefont{D.~A.~R.} \bibnamefont{Dalvit}},
  \bibnamefont{and} \bibinfo{author}{\bibfnamefont{P.~W.}
  \bibnamefont{Milonni}}, \bibinfo{journal}{{P}hys. {R}ev. {L}ett.}
  \textbf{\bibinfo{volume}{100}}, \bibinfo{pages}{183602}
  (\bibinfo{year}{2008}).

\bibitem[{\citenamefont{Yang et~al.}(2008)\citenamefont{Yang, Zeng, Xu, and
  Liu}}]{Yang2008}
\bibinfo{author}{\bibfnamefont{Y.}~\bibnamefont{Yang}},
  \bibinfo{author}{\bibfnamefont{R.}~\bibnamefont{Zeng}},
  \bibinfo{author}{\bibfnamefont{J.}~\bibnamefont{Xu}}, \bibnamefont{and}
  \bibinfo{author}{\bibfnamefont{S.}~\bibnamefont{Liu}},
  \bibinfo{journal}{{P}hys. {R}ev. {A}} \textbf{\bibinfo{volume}{77}},
  \bibinfo{pages}{015803} (\bibinfo{year}{2008}).

\bibitem[{\citenamefont{Henkel and Joulain}(2005)}]{Henkel2005}
\bibinfo{author}{\bibfnamefont{C.}~\bibnamefont{Henkel}} \bibnamefont{and}
  \bibinfo{author}{\bibfnamefont{K.}~\bibnamefont{Joulain}},
  \bibinfo{journal}{Europhys. {L}ett.} \textbf{\bibinfo{volume}{72}},
  \bibinfo{pages}{929} (\bibinfo{year}{2005}).

\bibitem[{\citenamefont{Toma\v{s}}(2005{\natexlab{a}})}]{Tomas2005}
\bibinfo{author}{\bibfnamefont{M.~S.} \bibnamefont{Toma\v{s}}},
  \bibinfo{journal}{Phys. {L}ett. {A}} \textbf{\bibinfo{volume}{342}},
  \bibinfo{pages}{381} (\bibinfo{year}{2005}{\natexlab{a}}).

\bibitem[{\citenamefont{Power and
  Thirunamachandran}(1995{\natexlab{a}})}]{Power1995a}
\bibinfo{author}{\bibfnamefont{E.~A.} \bibnamefont{Power}} \bibnamefont{and}
  \bibinfo{author}{\bibfnamefont{T.}~\bibnamefont{Thirunamachandran}},
  \bibinfo{journal}{{P}hys. {R}ev. {A}} \textbf{\bibinfo{volume}{51}},
  \bibinfo{pages}{3660} (\bibinfo{year}{1995}{\natexlab{a}}).

\bibitem[{\citenamefont{Rizzuto et~al.}(2004)\citenamefont{Rizzuto, Passant,
  and Persico}}]{Rizzuto2004}
\bibinfo{author}{\bibfnamefont{L.}~\bibnamefont{Rizzuto}},
  \bibinfo{author}{\bibfnamefont{R.}~\bibnamefont{Passant}}, \bibnamefont{and}
  \bibinfo{author}{\bibfnamefont{F.}~\bibnamefont{Persico}},
  \bibinfo{journal}{{P}hys. {R}ev. {A}} \textbf{\bibinfo{volume}{70}},
  \bibinfo{pages}{012107} (\bibinfo{year}{2004}).

\bibitem[{\citenamefont{Sherkunov}(2005)}]{Sherkunov2005}
\bibinfo{author}{\bibfnamefont{Y.}~\bibnamefont{Sherkunov}},
  \bibinfo{journal}{{P}hys. {R}ev. {A}} \textbf{\bibinfo{volume}{72}},
  \bibinfo{pages}{052703} (\bibinfo{year}{2005}).

\bibitem[{\citenamefont{Power and
  Thirunamachandran}(1995{\natexlab{b}})}]{Power1995}
\bibinfo{author}{\bibfnamefont{E.}~\bibnamefont{Power}} \bibnamefont{and}
  \bibinfo{author}{\bibfnamefont{T.}~\bibnamefont{Thirunamachandran}},
  \bibinfo{journal}{{C}hem. {P}hys.} \textbf{\bibinfo{volume}{198}},
  \bibinfo{pages}{5} (\bibinfo{year}{1995}{\natexlab{b}}).

\bibitem[{\citenamefont{Passante et~al.}(2005)\citenamefont{Passante, Persico,
  and Rizzuto}}]{Passante2005}
\bibinfo{author}{\bibfnamefont{R.}~\bibnamefont{Passante}},
  \bibinfo{author}{\bibfnamefont{F.}~\bibnamefont{Persico}}, \bibnamefont{and}
  \bibinfo{author}{\bibfnamefont{L.}~\bibnamefont{Rizzuto}},
  \bibinfo{journal}{{J}. of {M}od. {O}pt.} \textbf{\bibinfo{volume}{52}},
  \bibinfo{pages}{1957} (\bibinfo{year}{2005}).

\bibitem[{\citenamefont{Sherkunov}(2007{\natexlab{a}})}]{Sherkunov2007a}
\bibinfo{author}{\bibfnamefont{Y.}~\bibnamefont{Sherkunov}},
  \bibinfo{journal}{{J}. of {P}hys. {D}: {A}pp. {P}hys.}
  \textbf{\bibinfo{volume}{40}}, \bibinfo{pages}{86}
  (\bibinfo{year}{2007}{\natexlab{a}}).

\bibitem[{\citenamefont{Wylie and Sipe}(1984)}]{Wylie1984}
\bibinfo{author}{\bibfnamefont{J.~M.} \bibnamefont{Wylie}} \bibnamefont{and}
  \bibinfo{author}{\bibfnamefont{J.~E.} \bibnamefont{Sipe}},
  \bibinfo{journal}{Phys. {R}ev. {A}} \textbf{\bibinfo{volume}{30}},
  \bibinfo{pages}{185} (\bibinfo{year}{1984}).

\bibitem[{\citenamefont{Buhmann
  et~al.}(2004{\natexlab{a}})\citenamefont{Buhmann, Kn\"{o}ll, Welsch, and
  Dung}}]{Buhmann2004}
\bibinfo{author}{\bibfnamefont{S.~Y.} \bibnamefont{Buhmann}},
  \bibinfo{author}{\bibfnamefont{L.}~\bibnamefont{Kn\"{o}ll}},
  \bibinfo{author}{\bibfnamefont{D.-G.} \bibnamefont{Welsch}},
  \bibnamefont{and} \bibinfo{author}{\bibfnamefont{Ho Trung}
\bibnamefont{Dung}},
  \bibinfo{journal}{Phys. Rev. A} \textbf{\bibinfo{volume}{70}},
  \bibinfo{pages}{052117} (\bibinfo{year}{2004}{\natexlab{a}}).

\bibitem[{\citenamefont{Sherkunov}(2007{\natexlab{b}})}]{Sherkunov2007}
\bibinfo{author}{\bibfnamefont{Y.}~\bibnamefont{Sherkunov}},
  \bibinfo{journal}{{P}hy. {R}ev. {A}} \textbf{\bibinfo{volume}{75}},
  \bibinfo{pages}{012705} (\bibinfo{year}{2007}{\natexlab{b}}).

\bibitem[{\citenamefont{Sambale
  et~al.}(2008{\natexlab{a}})\citenamefont{Sambale, Welsch, Ho, and
  Buhmann}}]{Sambale2008a}
\bibinfo{author}{\bibfnamefont{A.}~\bibnamefont{Sambale}},
  \bibinfo{author}{\bibfnamefont{D.-G.} \bibnamefont{Welsch}},
  \bibinfo{author}{\bibfnamefont{Ho Trung} \bibnamefont{Dung}},
\bibnamefont{and}
  \bibinfo{author}{\bibfnamefont{S.~Y.} \bibnamefont{Buhmann}},
  \bibinfo{journal}{arxiv 0711.3369, submitted to {P}hys. {R}ev. {A}}
  (\bibinfo{year}{2008}{\natexlab{a}}).

\bibitem[{\citenamefont{Sambale
  et~al.}(2008{\natexlab{b}})\citenamefont{Sambale, Buhmann, Dung, and
  Welsch}}]{Sambale2008b}
\bibinfo{author}{\bibfnamefont{A.}~\bibnamefont{Sambale}},
  \bibinfo{author}{\bibfnamefont{S.~Y.} \bibnamefont{Buhmann}},
  \bibinfo{author}{\bibfnamefont{Ho Trung} \bibnamefont{Dung}},
\bibnamefont{and}
  \bibinfo{author}{\bibfnamefont{D.-G.} \bibnamefont{Welsch}},
  \bibinfo{journal}{arXiv 0809.3086, to be published in {P}hys. {S}cripta}
  (\bibinfo{year}{2008}{\natexlab{b}}).

\bibitem[{\citenamefont{Sandoghdar et~al.}(1992)\citenamefont{Sandoghdar,
  Sukenik, Hinds, and Haroche}}]{Sandoghdar1992}
\bibinfo{author}{\bibfnamefont{V.}~\bibnamefont{Sandoghdar}},
  \bibinfo{author}{\bibfnamefont{C.~I.} \bibnamefont{Sukenik}},
  \bibinfo{author}{\bibfnamefont{E.~A.} \bibnamefont{Hinds}}, \bibnamefont{and}
  \bibinfo{author}{\bibfnamefont{S.}~\bibnamefont{Ha\-roche}},
  \bibinfo{journal}{Phys. {R}ev. {L}ett.} \textbf{\bibinfo{volume}{68}},
  \bibinfo{pages}{3432} (\bibinfo{year}{1992}).

\bibitem[{\citenamefont{Chevrollier et~al.}(2001)\citenamefont{Chevrollier,
  Oria, de~Souza, Bloch, Fichet, and Ducloy}}]{Chevrollier2001}
\bibinfo{author}{\bibfnamefont{M.}~\bibnamefont{Chevrollier}},
  \bibinfo{author}{\bibfnamefont{M.}~\bibnamefont{Oria}},
  \bibinfo{author}{\bibfnamefont{J.~G.} \bibnamefont{de~Souza}},
  \bibinfo{author}{\bibfnamefont{D.}~\bibnamefont{Bloch}},
  \bibinfo{author}{\bibfnamefont{M.}~\bibnamefont{Fichet}}, \bibnamefont{and}
  \bibinfo{author}{\bibfnamefont{M.}~\bibnamefont{Ducloy}},
  \bibinfo{journal}{Phys. Rev. E} \textbf{\bibinfo{volume}{63}},
  \bibinfo{pages}{046610} (\bibinfo{year}{2001}).

\bibitem[{\citenamefont{Failache et~al.}(2003)\citenamefont{Failache, Altiel,
  Fichet, Bloch, and Ducloy}}]{Failache2003}
\bibinfo{author}{\bibfnamefont{H.}~\bibnamefont{Failache}},
  \bibinfo{author}{\bibfnamefont{S.}~\bibnamefont{Altiel}},
  \bibinfo{author}{\bibfnamefont{M.}~\bibnamefont{Fichet}},
  \bibinfo{author}{\bibfnamefont{D.}~\bibnamefont{Bloch}}, \bibnamefont{and}
  \bibinfo{author}{\bibfnamefont{M.}~\bibnamefont{Du\-cloy}},
  \bibinfo{journal}{Eur. Phys. J. D} \textbf{\bibinfo{volume}{23}},
  \bibinfo{pages}{237} (\bibinfo{year}{2003}).

\bibitem[{\citenamefont{Ninham and Daicic}(1998)}]{Ninham1998}
\bibinfo{author}{\bibfnamefont{B.}~\bibnamefont{Ninham}} \bibnamefont{and}
  \bibinfo{author}{\bibfnamefont{J.}~\bibnamefont{Daicic}},
  \bibinfo{journal}{Phys. Rev. A} \textbf{\bibinfo{volume}{57}},
  \bibinfo{pages}{1870} (\bibinfo{year}{1998}).

\bibitem[{\citenamefont{Mostepanenko et~al.}(2006)\citenamefont{Mostepanenko,
  Bezerra, Decca, Geyer, Fischbach, Klimchitskaya, Krause, and
  L\`{o}pez}}]{Mostepanenko2006}
\bibinfo{author}{\bibfnamefont{V.~M.} \bibnamefont{Mostepanenko}},
  \bibinfo{author}{\bibfnamefont{V.~B.} \bibnamefont{Bezerra}},
  \bibinfo{author}{\bibfnamefont{R.~S.} \bibnamefont{Decca}},
  \bibinfo{author}{\bibfnamefont{B.}~\bibnamefont{Geyer}},
  \bibinfo{author}{\bibfnamefont{E.}~\bibnamefont{Fischbach}},
  \bibinfo{author}{\bibfnamefont{G.~L.} \bibnamefont{Klimchitskaya}},
  \bibinfo{author}{\bibfnamefont{D.~E.} \bibnamefont{Krause}},
  \bibnamefont{and}
  \bibinfo{author}{\bibfnamefont{D.}~\bibnamefont{L\`{o}pez}},
  \bibinfo{journal}{{J}. of {P}hys. {A}: {M}athematical and {G}eneral}
  \textbf{\bibinfo{volume}{39}}, \bibinfo{pages}{6589} (\bibinfo{year}{2006}).

\bibitem[{\citenamefont{Decca et~al.}(2005)\citenamefont{Decca, L\'{o}pez,
  Fischbach, Klimchitskaya, Krause, and Mostepanenko}}]{Decca2005}
\bibinfo{author}{\bibfnamefont{R.~S.} \bibnamefont{Decca}},
  \bibinfo{author}{\bibfnamefont{D.}~\bibnamefont{L\'{o}pez}},
  \bibinfo{author}{\bibfnamefont{E.}~\bibnamefont{Fischbach}},
  \bibinfo{author}{\bibfnamefont{G.~L.} \bibnamefont{Klimchitskaya}},
  \bibinfo{author}{\bibfnamefont{D.~E.} \bibnamefont{Krause}},
  \bibnamefont{and}
  \bibinfo{author}{\bibfnamefont{V.}~\bibnamefont{Mostepanenko}},
  \bibinfo{journal}{{A}nn. {P}hys. {NY}} \textbf{\bibinfo{volume}{318}},
  \bibinfo{pages}{37} (\bibinfo{year}{2005}).

\bibitem[{\citenamefont{Klimchitskaya et~al.}(2005)\citenamefont{Klimchitskaya,
  Decca, Fischbach, Krause, L\'{o}pez, and Mostepanenko}}]{Klimchitskaya2005}
\bibinfo{author}{\bibfnamefont{G.~L.} \bibnamefont{Klimchitskaya}},
  \bibinfo{author}{\bibfnamefont{R.~S.} \bibnamefont{Decca}},
  \bibinfo{author}{\bibfnamefont{E.}~\bibnamefont{Fischbach}},
  \bibinfo{author}{\bibfnamefont{D.~E.} \bibnamefont{Krause}},
  \bibinfo{author}{\bibfnamefont{D.}~\bibnamefont{L\'{o}pez}},
  \bibnamefont{and}
  \bibinfo{author}{\bibfnamefont{V.}~\bibnamefont{Mostepanenko}},
  \bibinfo{journal}{{I}nt. {J}. {M}od. {P}hys. {A}}
  \textbf{\bibinfo{volume}{20}}, \bibinfo{pages}{2205} (\bibinfo{year}{2005}).

\bibitem[{\citenamefont{McLachlan}(1963)}]{McLachlan1963}
\bibinfo{author}{\bibfnamefont{A.}~\bibnamefont{McLachlan}},
  \bibinfo{journal}{{P}roc. {R}. {S}oc. {L}ond. {S}er. {A}}
  \textbf{\bibinfo{volume}{274}} (\bibinfo{year}{1963}).

\bibitem[{\citenamefont{Henkel et~al.}(2002)\citenamefont{Henkel, Joulain,
  Mulet, and Greffet}}]{Henkel2002}
\bibinfo{author}{\bibfnamefont{C.}~\bibnamefont{Henkel}},
  \bibinfo{author}{\bibfnamefont{K.}~\bibnamefont{Joulain}},
  \bibinfo{author}{\bibfnamefont{J.-P.} \bibnamefont{Mulet}}, \bibnamefont{and}
  \bibinfo{author}{\bibfnamefont{J.-J.} \bibnamefont{Greffet}},
  \bibinfo{journal}{J. {O}pt. {A}: {P}ure {A}ppl. {O}pt.}
  \textbf{\bibinfo{volume}{4}}, \bibinfo{pages}{S109} (\bibinfo{year}{2002}).

\bibitem[{\citenamefont{Gorza and Ducloy}(2006)}]{Gorza2006}
\bibinfo{author}{\bibfnamefont{M.-P.} \bibnamefont{Gorza}} \bibnamefont{and}
  \bibinfo{author}{\bibfnamefont{M.}~\bibnamefont{Ducloy}},
  \bibinfo{journal}{{E}ur. {P}hys. {J}. D} \textbf{\bibinfo{volume}{40}},
  \bibinfo{pages}{343} (\bibinfo{year}{2006}).

\bibitem[{\citenamefont{Antezza et~al.}(2005)\citenamefont{Antezza, Pitaevskii,
  and Stringari}}]{Antezza2005}
\bibinfo{author}{\bibfnamefont{M.}~\bibnamefont{Antezza}},
  \bibinfo{author}{\bibfnamefont{L.~P.} \bibnamefont{Pitaevskii}},
  \bibnamefont{and}
  \bibinfo{author}{\bibfnamefont{S.}~\bibnamefont{Stringari}},
  \bibinfo{journal}{{P}hys. {R}ev. {L}ett.} \textbf{\bibinfo{volume}{95}},
  \bibinfo{pages}{113202} (\bibinfo{year}{2005}).

\bibitem[{\citenamefont{Harber et~al.}(2005)\citenamefont{Harber, Obrecht,
  McGuirk, and Cornell}}]{Harber2005}
\bibinfo{author}{\bibfnamefont{D.~M.} \bibnamefont{Harber}},
  \bibinfo{author}{\bibfnamefont{J.~M.} \bibnamefont{Obrecht}},
  \bibinfo{author}{\bibfnamefont{J.~M.} \bibnamefont{McGuirk}},
  \bibnamefont{and} \bibinfo{author}{\bibfnamefont{E.~A.}
  \bibnamefont{Cornell}}, \bibinfo{journal}{Phys. Rev. A}
  \textbf{\bibinfo{volume}{72}}, \bibinfo{pages}{033610}
  (\bibinfo{year}{2005}).

\bibitem[{\citenamefont{Obrecht et~al.}(2007)\citenamefont{Obrecht, Wild,
  Antezza, Pitaevskii, Stringari, and Cornell}}]{Obrecht2007}
\bibinfo{author}{\bibfnamefont{J.~M.} \bibnamefont{Obrecht}},
  \bibinfo{author}{\bibfnamefont{R.~J.} \bibnamefont{Wild}},
  \bibinfo{author}{\bibfnamefont{M.}~\bibnamefont{Antezza}},
  \bibinfo{author}{\bibfnamefont{L.~P.} \bibnamefont{Pitaevskii}},
  \bibinfo{author}{\bibfnamefont{S.}~\bibnamefont{Stringari}},
  \bibnamefont{and} \bibinfo{author}{\bibfnamefont{E.~A.}
  \bibnamefont{Cornell}}, \bibinfo{journal}{{P}hys. {R}ev. {L}ett.}
  \textbf{\bibinfo{volume}{98}}, \bibinfo{pages}{063201}
  (\bibinfo{year}{2007}).

\bibitem[{\citenamefont{Buhmann and Scheel}(2008{\natexlab{a}})}]{Buhmann2008b}
\bibinfo{author}{\bibfnamefont{S.~Y.} \bibnamefont{Buhmann}} \bibnamefont{and}
  \bibinfo{author}{\bibfnamefont{S.}~\bibnamefont{Scheel}},
  \bibinfo{journal}{arxiv: 0803.0738}  (\bibinfo{year}{2008}{\natexlab{a}}).

\bibitem[{\citenamefont{Sherkunov}(2008)}]{Sherkunov2008}
\bibinfo{author}{\bibfnamefont{Y.}~\bibnamefont{Sherkunov}},
  \bibinfo{journal}{arxiv: 0608.2620}  (\bibinfo{year}{2008}).

\bibitem[{\citenamefont{Scully and Zubairy}(1997)}]{Scully1997}
\bibinfo{author}{\bibfnamefont{M.}~\bibnamefont{Scully}} \bibnamefont{and}
  \bibinfo{author}{\bibfnamefont{M.~S.} \bibnamefont{Zubairy}},
  \emph{\bibinfo{title}{Quantum optics}} (\bibinfo{publisher}{{C}ambr. {U}niv.
  {P}ress}, \bibinfo{year}{1997}).

\bibitem[{\citenamefont{Popov and Shalaev}(2006)}]{Popov2006}
\bibinfo{author}{\bibfnamefont{A.~K.} \bibnamefont{Popov}} \bibnamefont{and}
  \bibinfo{author}{\bibfnamefont{V.~M.} \bibnamefont{Shalaev}},
  \bibinfo{journal}{{O}pt. {L}ett.} \textbf{\bibinfo{volume}{31}},
  \bibinfo{pages}{2169} (\bibinfo{year}{2006}).

\bibitem[{\citenamefont{Stockmann}(2007)}]{Stockmann2007}
\bibinfo{author}{\bibfnamefont{M.}~\bibnamefont{Stockmann}},
  \bibinfo{journal}{Phys. Rev. Lett.} \textbf{\bibinfo{volume}{98}},
  \bibinfo{pages}{177404} (\bibinfo{year}{2007}).

\bibitem[{\citenamefont{Leonhardt and Philbin}(2007)}]{Leonhardt2007}
\bibinfo{author}{\bibfnamefont{U.}~\bibnamefont{Leonhardt}} \bibnamefont{and}
  \bibinfo{author}{\bibfnamefont{T.~G.} \bibnamefont{Philbin}},
  \bibinfo{journal}{{N}. {J}. of {P}hys.} \textbf{\bibinfo{volume}{9}},
  \bibinfo{pages}{254} (\bibinfo{year}{2007}).

\bibitem[{\citenamefont{Zhao}(2003)}]{Zhao2003}
\bibinfo{author}{\bibfnamefont{Y.~P.} \bibnamefont{Zhao}},
  \bibinfo{journal}{{A}cta {M}echanica {S}inica} \textbf{\bibinfo{volume}{19}},
  \bibinfo{pages}{1} (\bibinfo{year}{2003}).

\bibitem[{\citenamefont{Raabe and Welsch}(2008)}]{Raabe2008}
\bibinfo{author}{\bibfnamefont{C.}~\bibnamefont{Raabe}} \bibnamefont{and}
  \bibinfo{author}{\bibfnamefont{D.-G.} \bibnamefont{Welsch}},
  \bibinfo{journal}{The Eur. Phys. J. - Special Topics}
  \textbf{\bibinfo{volume}{160}}, \bibinfo{pages}{371} (\bibinfo{year}{2008}).

\bibitem[{\citenamefont{Jeffers et~al.}(1996)\citenamefont{Jeffers, Barnett,
  Loudon, Matloob, and Artoni}}]{Jeffers1996}
\bibinfo{author}{\bibfnamefont{J.}~\bibnamefont{Jeffers}},
  \bibinfo{author}{\bibfnamefont{S.~M.} \bibnamefont{Barnett}},
  \bibinfo{author}{\bibfnamefont{R.}~\bibnamefont{Loudon}},
  \bibinfo{author}{\bibfnamefont{R.}~\bibnamefont{Matloob}}, \bibnamefont{and}
  \bibinfo{author}{\bibfnamefont{M.}~\bibnamefont{Artoni}},
  \bibinfo{journal}{{O}pt. {C}ommun.} \textbf{\bibinfo{volume}{131}},
  \bibinfo{pages}{66} (\bibinfo{year}{1996}).

\bibitem[{\citenamefont{Matloob et~al.}(1997)\citenamefont{Matloob, Loudon,
  Artoni, Barnett, and Jeffers}}]{Matloob1997}
\bibinfo{author}{\bibfnamefont{R.}~\bibnamefont{Matloob}},
  \bibinfo{author}{\bibfnamefont{R.}~\bibnamefont{Loudon}},
  \bibinfo{author}{\bibfnamefont{M.}~\bibnamefont{Artoni}},
  \bibinfo{author}{\bibfnamefont{S.~M.} \bibnamefont{Barnett}},
  \bibnamefont{and} \bibinfo{author}{\bibfnamefont{J.}~\bibnamefont{Jeffers}},
  \bibinfo{journal}{{P}hys. {R}ev. {A}} \textbf{\bibinfo{volume}{55}},
  \bibinfo{pages}{1623} (\bibinfo{year}{1997}).

\bibitem[{\citenamefont{Kubo et~al.}(1998)\citenamefont{Kubo, Toda, and
  Hashitsume}}]{Kubo1998}
\bibinfo{author}{\bibfnamefont{R.}~\bibnamefont{Kubo}},
  \bibinfo{author}{\bibfnamefont{M.}~\bibnamefont{Toda}}, \bibnamefont{and}
  \bibinfo{author}{\bibfnamefont{N.}~\bibnamefont{Hashitsume}},
  \emph{\bibinfo{title}{Statis\-tical Physics II: Nonequilibrium {S}tatistical
  {M}echanics}} (\bibinfo{publisher}{Springer Verlag Berlin},
  \bibinfo{year}{1998}).

\bibitem[{\citenamefont{Melrose and McPhedran}(2003)}]{Melrose2003}
\bibinfo{author}{\bibfnamefont{D.~B.} \bibnamefont{Melrose}} \bibnamefont{and}
  \bibinfo{author}{\bibfnamefont{R.~C.} \bibnamefont{McPhedran}},
  \emph{\bibinfo{title}{Electromagnetic processes in dispersive media : {A}
  treatment based on the dielectric tensor}} (\bibinfo{publisher}{{C}ambridge
  {U}niv. {P}ress,}, \bibinfo{year}{2003}).

\bibitem[{\citenamefont{Scheel et~al.}(1998)\citenamefont{Scheel, Kn\"{o}ll,
  and Welsch}}]{Scheel1998}
\bibinfo{author}{\bibfnamefont{S.}~\bibnamefont{Scheel}},
  \bibinfo{author}{\bibfnamefont{L.}~\bibnamefont{Kn\"{o}ll}},
  \bibnamefont{and} \bibinfo{author}{\bibfnamefont{D.-G.}
  \bibnamefont{Welsch}}, \bibinfo{journal}{Phys. Rev. A}
  \textbf{\bibinfo{volume}{58}}, \bibinfo{pages}{700} (\bibinfo{year}{1998}).

\bibitem[{\citenamefont{Raabe and Welsch}(2006)}]{Raabe2006}
\bibinfo{author}{\bibfnamefont{C.}~\bibnamefont{Raabe}} \bibnamefont{and}
  \bibinfo{author}{\bibfnamefont{D.-G.} \bibnamefont{Welsch}},
  \bibinfo{journal}{Phys. {R}ev. {A}} \textbf{\bibinfo{volume}{73}},
  \bibinfo{pages}{063822} (\bibinfo{year}{2006}), \bibinfo{note}{note erratum
  Phys. Rev. A 74 (1) (2006) 019901(E) .}

\bibitem[{\citenamefont{Buhmann
  et~al.}(2004{\natexlab{b}})\citenamefont{Buhmann, Ho, and
  Welsch}}]{Buhmann2004pertur}
\bibinfo{author}{\bibfnamefont{S.~Y.} \bibnamefont{Buhmann}},
  \bibinfo{author}{\bibfnamefont{Ho Trung} \bibnamefont{Dung}},
\bibnamefont{and}
  \bibinfo{author}{\bibfnamefont{D.-G.} \bibnamefont{Welsch}},
  \bibinfo{journal}{J. Opt. B: Quantum and Semiclassical Optics}
  \textbf{\bibinfo{volume}{6}}, \bibinfo{pages}{127}
  (\bibinfo{year}{2004}{\natexlab{b}}).

\bibitem[{\citenamefont{Buhmann and Scheel}(2008{\natexlab{b}})}]{Buhmann2008a}
\bibinfo{author}{\bibfnamefont{S.~Y.} \bibnamefont{Buhmann}} \bibnamefont{and}
  \bibinfo{author}{\bibfnamefont{S.}~\bibnamefont{Scheel}},
  \bibinfo{journal}{arxiv 0806.2211}  (\bibinfo{year}{2008}{\natexlab{b}}).

\bibitem[{\citenamefont{Toma\v{s}}(2005{\natexlab{b}})}]{Tomas2005b}
\bibinfo{author}{\bibfnamefont{M.}~\bibnamefont{Toma\v{s}}},
  \bibinfo{journal}{{P}hys. {R}ev. {A}} \textbf{\bibinfo{volume}{71}},
  \bibinfo{pages}{060101(R)} (\bibinfo{year}{2005}{\natexlab{b}}).

\bibitem[{\citenamefont{Buhmann et~al.}(2006)\citenamefont{Buhmann, Safari,
  Welsch, and Ho}}]{Buhmann2006a}
\bibinfo{author}{\bibfnamefont{S.~Y.} \bibnamefont{Buhmann}},
  \bibinfo{author}{\bibfnamefont{H.}~\bibnamefont{Safari}},
  \bibinfo{author}{\bibfnamefont{D.-G.} \bibnamefont{Welsch}},
  \bibnamefont{and} \bibinfo{author}{\bibfnamefont{Ho Trung}
\bibnamefont{Dung}},
  \bibinfo{journal}{{O}pen {S}ys. \& {I}nformation {D}yn.}
  \textbf{\bibinfo{volume}{13}}, \bibinfo{pages}{427} (\bibinfo{year}{2006}).

\bibitem[{\citenamefont{Buhmann and Welsch}(2006)}]{Buhmann2006b}
\bibinfo{author}{\bibfnamefont{S.~Y.} \bibnamefont{Buhmann}} \bibnamefont{and}
  \bibinfo{author}{\bibfnamefont{D.-G.} \bibnamefont{Welsch}},
  \bibinfo{journal}{Appl. {P}hys. {B}} \textbf{\bibinfo{volume}{82}},
  \bibinfo{pages}{189} (\bibinfo{year}{2006}).

\bibitem[{\citenamefont{Toma\v{s}}(1995)}]{Tomas1995}
\bibinfo{author}{\bibfnamefont{M.~S.} \bibnamefont{Toma\v{s}}},
  \bibinfo{journal}{{P}hys. {R}ev. {A}} \textbf{\bibinfo{volume}{51}},
  \bibinfo{pages}{2545} (\bibinfo{year}{1995}).

\end{thebibliography}

\end{document}